\documentstyle[aps,prl,twocolumn]{revtex}
\begin{document}
\author{Yong-Jun Liu$^{1,2}$ and Chang-De Gong$^{3,1}$}
\title{Quantum Phase Transitions in Spin-$\frac 12$ Frustrated Molecular Cluster:
the Numerical Evidence}
\address{$^1$National Key Laboratory of Solid States of Microstructure, Nanjing\\
University, Nanjing 210093, PRC\\
$^2$Complexity Science Center, Yangzhou University, Yangzhou 225002, PRC\\
$^3$CCAST (World Laboratory), P.O. Box 8730, Beijing 100080, PRC}
\maketitle

\begin{abstract}
By exact diagonalization, we investigate several spin-$\frac 12$ $J_1-J_2$
real clusters of molecular scale with different shapes. Our calculations
show that when the ratio, $\eta $, of next nearest neighbor to nearest
neighbor bonds is equal to $1$, even the cluster of only $25$ sites exhibits
the bulk behaviors and has the quantum phase transitions. Two effective
critical points are around $J_2/J_1=0.3762$ and $0.612$ respectively. They
are very close to those of the infinite $J_1-J_2$ square lattice. But, when $%
\eta \leq 0.85$, the quantum phase transition around $J_2/J_1=0.3762$
disappears. By calculating the distributions of the average values of $S_i^z$%
, the bulk behaviors are demonstrated graphically. In the intermediate
phase, the sites on the corners have distinctly different character from the
other sites. The distribution is obviously centralized on the corner sites.
\end{abstract}


The two-dimensional spin-$\frac 12$ $J_1-J_2$ Heisenberg model has been the
object of intense investigation through years since the antiferromagnetism
can be destabilized by frustrations and it exhibits rich physics and diverse
properties. Its Hamiltonian reads

\begin{equation}
H=J_1\sum_{\left\langle n.n.\right\rangle }\vec{S}_i\cdot \vec{S}%
_j+J_2\sum_{\left\langle n.n.n.\right\rangle }\vec{S}_i\cdot \vec{S}_j,
\label{Hami}
\end{equation}
where $\left\langle n.n.\right\rangle $ and $\left\langle
n.n.n.\right\rangle $ are the sum over the nearest neighbors and next
nearest neighbors respectively, $J_1>0$ and $J_2\geq 0$. For the
two-dimensional antiferromagnet without frustration $(J_2=0)$,
notwithstanding a mathematically rigorous solution is lacking, it is by now
well established that the ground state (GS) has N\'{e}el order by many works 
\cite{Gelfand2,Reger}. When $J_2>0$, the competition of purely magnetic
interactions can lead to the destruction of the long-range order. Many
analytical approximations \cite
{Gelfand,Chandra,Chubukov,Kotov,Gelfand2,Read,Kotov2,Singh,Sachdev} and
numerical techniques \cite{Johnston,Sandvik,Dagotto,Schulz} are applied to
investigate the phase diagram of the $J_1-J_2$ model. For small $J_2/J_1$,
the GS will keep N\'{e}el order described by a wave vector $(\pi ,\pi )$
since the frustrations are too weak to destroy it. When $J_2/J_1$ is large
enough, the GS is dominated by the $n.n.n.$ interactions and has a certain
type of collinear order described by $(\pi ,0)$ or $(0,\pi )$. But, for
intermediate $J_2/J_1$, the competition between $J_1$ and $J_2$ terms
results in a spin-liquid GS with a gap to magnetic excitations. The two
critical points of phase transition are at $J_2/J_1=0.38$ and $0.60$
respectively \cite{Gelfand2,Read,Kotov2,Singh}.

The situation of real cluster, when it has small size, may be different from
that of infinite system (the word `real' means that the free boundary
conditions are taken). Especially for real clusters of molecular scale ,
which has only tens of sites, the boundary effect might be heavy since many
sites are on its boundaries. For instance, the $4\times 4$, $5\times 5$ and $%
6\times 6$ real clusters have $75\%,64\%$ and $56\%$ sites on their
boundaries respectively. In order to understand the spin status of real
clusters, we introduce

\begin{equation}
Q(\vec{q})=\frac 1{N^2}\sum_{(i,j)}\left\langle G\left| \vec{S}_i\cdot \vec{S%
}_j\right| G\right\rangle \exp (i\vec{q}\cdot (\vec{r}_i-\vec{r}_j)),
\label{r-space}
\end{equation}
as a measure of magnetic orders, where $N$ is the total number of sites, $%
\vec{r}_i$ and $\vec{r}_j$ the coordinates of sites $i$ and $j$
respectively, $\left| G\right\rangle $ represents the GS. At first, we
calculate $Q(\pi ,\pi )$ of the $5\times 5$ real cluster by exact
diagonalization. Our calculation shows that $Q(\pi ,\pi )$ decays
continuously and monotonically in the region of $J_2/J_1\sim 0-1.0$ (curve
(e) in Fig. 2). Especially, unlike the case of the infinite system, the real
cluster has no quantum phase transition around $J_2/J_1=0.38$. One may think
that the cause is too small freedom to form the bulk behavior of spins. It
is meaningful to investigate whether the real clusters of molecular scale
always keep this property. However, the real cluster with other shape might
have different properties due to the boundary effects. If define $\eta
\equiv N_{nnn}^{bond}/N_{nn}^{bond}$, where $N_{nn}^{bond}$ and $%
N_{nnn}^{bond}$ represent the numbers of $n.n.$ and $n.n.n.$ bonds, we
notice that $\eta =1$ for the infinite $J_1-J_2$ square lattice. $\eta $ may
be an important factor for spin status. In other words, if a real cluster
posses a structure of $\eta =1$, it has possibly the similar quantum
transition as the infinite system. When its boundaries are set up along
diagonals, the real cluster is able to hold $\eta =1$. Calculating $Q(\pi
,\pi )$ of cluster (a) (Fig. 1), we find that it is quite different from the
real $5\times 5$ cluster and $Q(\pi ,\pi )$ has a step-like change around $%
J_2/J_1=0.3762$ (Fig. 2). In a very small interval of $J_2/J_1$, which is
less than $4\times 10^{-4}$, $Q(\pi ,\pi )$ falls $55\%$. The abrupt change
in such a narrow parameter region means that the small real cluster has
possibly the bulk behaviors. {\it i.e.}, there exists a quantum phase
transition around $J_2/J_1=0.3762$. Larger $Q(\pi ,\pi )$ for $J_2/J_1\leq
0.3760$ means the N\'{e}el-like order \cite{Order}, whereas small $Q(\pi
,\pi )$ for $J_2/J_1\geq 0.3764$ means its losing. We will give the further
evidence for this conclusion. Unlike the infinite system has $Q(\pi ,\pi )=0$
when the AF long-range order vanishes, here zero $Q(\pi ,\pi )$ can not be
achieved since the real cluster is finite. In the region $0.3764\leq
J_2/J_1\leq 1.0$, although $Q(\pi ,\pi )$ decays monotonically as $J_2/J_1$
increasing, it keeps positive. The real frustrated cluster always has a
little antiferromagnetism.

Generally, the bulk behavior occurs for these clusters with size large
enough. The too small size may lead to its disappearance due to the
limitation of freedom. But, our calculations show that when the real cluster
of molecular scale keeps $\eta =1$, it has the bulk properties, especially
the quantum phase transition around $J_2/J_1=0.38$. Even for the real
cluster of $13$ sites with the similar shape, there also exists a step-like
change of $Q(\pi ,\pi )$ within $0.3381<J_2/J_1<0.3886$. It is noticed that
cluster (e), {\it i.e.} the $5\times 5$ real cluster, has $\eta =0.80$.
Three other real clusters (b), (c) and (d) (Fig. 1), which have the same
number of sites but the different values of $\eta $ ($\eta =0.9$, $0.89$ and 
$0.85$ respectively), are investigated. As $\eta $ decreasing, the jump
around $J_2/J_1=0.38$ becomes smaller and smaller (Fig. 2). When $\eta \leq
0.85$, the step-like change disappears. Our calculations support the
conclusion that $\eta $ is an important factor for the quantum phase
transitions of real clusters and the enhance of $\eta $ advantages the bulk
behavior.

For the frustrated Heisenberg antiferromagnets, when the effect of $J_2$
terms exceeds that of $J_1$ terms, the $n.n.n.$ interactions may result in
the establishing of magnetic orders in its subsystems. If carefully checking
the $Q(\pi ,\pi )$ curves of these five clusters, one can find inflexions at
larger $J_2/J_1$ (Fig. 2). The difference of their geometric symmetries
causes the critical $J_2/J_1$ values obviously different since for such
small clusters, the symmetries may affect heavily the spin-Peierls states of
the disordered phase. One notices that same as the infinite square lattice,
cluster (a) keeps the rotation symmetry and $\eta =1$. Its singular point is
at $J_2/J_1=0.612(\pm 0.002)$. The slope of $Q(\pi ,\pi )$ increases when $%
J_2/J_1<0.612$ while deceases when $J_2/J_1>0.612$. For convenience, the
real cluster is divided into two subsystems $A$ and $B$ (Fig. 1). To
understand whether there is quantum transition around $J_2/J_1\approx 0.612$%
, it is useful to calculate $Q(\pi ,0)$ by formula (\ref{r-space}). As shown
in Fig. 2, $Q(\pi ,0)$ changes rapidly round $J_2/J_1\approx 0.612$, and it
keeps a obviously larger value for large $J_2/J_1$ than that for small $%
J_2/J_1$. It implies the existence of the N\'{e}el-like order in both of
subsystems $A$ and $B$ while no N\'{e}el-like order for the whole real
cluster. When $J_2/J_1<0.612$, the competition of $J_1$ and $J_2$ leads to
the disordered GS. Namely, the GS has the character of spin liquid.

The total spin of the GS of cluster (a) is nonzero since it has odd sites.
It provide us another way to investigate the spin status. The spin-spin
correlations lead to the probability of spin up unequal to that of spin down
on each site in the $M$ space, here $M$ is the magnetic quantum number.
Consequently $\overline{S_i^z}\equiv \left\langle G^{+}\left| S_i^z\right|
G^{+}\right\rangle \neq 0$, here $\left| G^{+}\right\rangle $ involves only
the ground states in the $M>0$ subspaces. (It is quite different from the
case of the GS with zero total spin, in which the spin up-down symmetry
makes $\overline{S_i^z}$ being zero exactly on each site.) The distribution
of $\overline{S_i^z}$ on sites depends on the spin-spin correlations. The
study of it is helpful to understand the spin status. Fig. 3 shows the cases
of cluster (a) for $J_2/J_1=0.37$, $0.40$, $0.60$ and $0.88$ respectively.
At $J_2/J_1=0.37$, the average value of $\overline{S_i^z}$ for the sites in
subsystem $A$ is $2.04\times 10^{-1}$ and its statistical fluctuation $%
5.48\times 10^{-3}$, and that for the sites in subsystem $B$ is $-1.41\times
10^{-1}$ and its fluctuation $1.43\times 10^{-3}$. {\it i.e.}, the
distribution of $\overline{S_i^z}$ is approximately uniform. It gives a good
picture of the N\'{e}el-like order. But, when $J_2/J_1\geq 0.3764$, the
distribution is rather different. Fig. 3(b) and 3(c) shows the cases of $%
J_2/J_1=0.40$ and $0.60$ respectively. For convenience to describe, we call
the sites on four corners as corner sites, and the other $21$ sites as the
inner sites (Fig. 1). At $J_2/J_1=0.40$ and $0.60$, $\overline{S_i^z}$ on
each corner site is equal to $1.88\times 10^{-1}$ and $1.00\times 10^{-1}$
respectively, whereas $\overline{S_i^z}$ on inner sites are distinctly small
and equal to $2.63\times 10^{-2}$ and $2.94\times 10^{-2}$ respectively. The
fact that the magnitudes of $\overline{S_i^z}$ on inner sites for $%
J_2/J_1\geq 0.3764$ are much smaller than those for $J_2/.J_1\leq 0.3760$
demonstrates the disappearance of the N\'{e}el-like order. For intermediate $%
J_2/J_1$, on each corner site, $\overline{S_i^z}$ always keeps positive, and
its magnitude is much larger than those of the inner sites. Such
distribution results from spin disorder (spin liquid) behavior. In this
case, the spins on corner sites are in a special status.

If $J_2/J_1$ increasing further, the $n.n.n.$ interactions lead to the
occurrence of a new ordered quantum phase. Fig. 3(d) shows the case of $%
J_2/J_1=0.88$. In subsystem $B$, the average value of positive $\overline{%
S_i^z}$ is $1.93\times 10^{-1}$, and the fluctuation $4.67\times 10^{-3}$.
The four corner sites with negative $\overline{S_i^z}$ have the same value $%
-1.37\times 10^{-1}$ due to the geometry of cluster (a). The distribution of 
$\overline{S_i^z}$ in subsystem $B$ exhibits the AF N\'{e}el-like order. In
subsystem $A$, $\overline{S_i^z}$ are very small and their average value is $%
5.11\times 10^{-3}$. It does not means that the spin status is disordered.
By contrary, it is also ordered by the calculation of $Q(\pi ,0)$. A simple
explanation to the smallness of $\overline{S_i^z}$ in subsystem $A$ is
given: cluster (a) consists of two subsystems $A$ and $B$ by $J_1$ coupling.
It is noticed that subsystem $B$ has odd sites, but subsystem $A$ has even
sites. If $J_1=0$, the distribution of $\overline{S_i^z}$ in subsystem $B$
can be used to show the N\'{e}el picture from $J_2$ interactions. But, in
subsystem $A$, we can not do it in the same way since $\overline{S_i^z}=0$
exactly. For large $J_2/J_1$, the $n.n.n.$ interactions dominate the physics
of the cluster. Consequently, the site in subsystem $A$ has small $\overline{%
S_i^z}$.

From our calculations, the following conclusions can be obtained. 1).
although it has only $25$ sites and nearly half of all sites on its
boundaries, the molecular cluster (a) exhibits the obvious bulk behaviors.
For small and large $J_2/J_1$, cluster (a) is in two different magnetic
ordered phases. For the intermediate $J_2/J_1$, the spin status is
disordered. Two kinds of quantum phase transitions take place around $%
J_2/J_1=0.3762$ and $0.612$. respectively. 2). for cluster (a) with
intermediate $J_2/J_1$, the spins on corner sites are in a special state.
And their magnitudes of $S_i^z$ are remarkably larger than those on inner
sites. We think that the larger real cluster with the similar shape as
cluster (a) will keep this character. 3). $\eta $ plays an important role
for real cluster to keep the bulk behaviors. Its increasing advantages the
formation of bulk behaviors. From our calculations, we think that only for
real clusters with $\eta $ near $1$, it is possible that there exists
quantum phase transition around $J_2/J_1=0.38$.

It is somewhat surprise that the two critical values $J_2/J_1=0.3762$ and $%
0.612$ of cluster (a) are very close to those of infinite square lattice $%
J_2/J_1=0.38$ and $0.60$ \cite{Sushkov}. The fact that the infinite $J_1-J_2$
square lattice has the same $\eta =1$ as cluster (a) may be responsible for
the result. $\eta $ is a key factor for real cluster to possess the similar
phase diagram as that in infinite system. By using exact diagonalization, H.
J. Schulz {\it et al.} studied the $6\times 6$ cluster under the periodic
boundary conditions. Their calculations show that there is no step-like
change of $Q(\pi ,\pi )$ around $J_2/J_1=0.38$, and $Q(\pi ,\pi )$ becomes
obviously small only when $J_2/J_1>0.6$ (see Fig. 3 (a) in Ref. \cite{Schulz}%
). To obtain reliable critical values of infinite system, one must resort to
scaling analyses like doing in Ref. \cite{Schulz}. In addition, the $6\times
6$ cluster has $56\%$ sites on its boundaries. Although cluster (a) has
remarkably fewer sites, the boundary effect might be weaker than the $%
6\times 6$ cluster since it only has $46\%$ sites on its boundaries. The
real clusters with such shape as cluster (a) might be better choice than the
conventional $n\times n$ cluster under periodic conditions to study some
properties of the infinite $J_1-J_2$ square lattice. But, the calculable
sizes of the clusters with such geometric shape are only $13$ and $25$. It
is difficult to obtain the results of infinite system by ordinary
extrapolation since the scarcity of available points. One notices that the
region $(0.3760\sim 0.3764)$ of $J_2/J_1$ of the real cluster of $25$ sites,
in which a step-like jump of $Q(\pi ,\pi )$ takes place, is within that of
the real cluster of $13$ sites $(0.3381\sim 0.3886)$. We think that the $%
J_2/J_1$ region of the larger real cluster will fall into $0.3760\sim 0.3764$%
. Then, we speculate that the critical value of the infinite system is $%
J_2/J_1=0.3762(\pm 0.0002)$. It may be more accurate than the previous
result $J_2/J_1=0.38$.

The part of calculations in this work have been done on the SGI Origin 2000
in the Group of Computational Condensed Matter Physics, National Laboratory
of Solid State Microstructures, Nanjing University. This work was partially
supported by the Ministry of Science and Technology of China under Grant No.
NKBRSF-G19990646.

\begin{figure}[tbp]
\caption{Five kinds of real clusters of 25 spins with different geometric
shapes. All the sites except (a) $1^{^{\prime }}$, $2^{^{\prime }}$, $%
3^{^{\prime }}$ and $4^{^{\prime }}$, (b) $1$, $2^{^{\prime }}$, $%
3^{^{\prime }}$ and $4^{^{\prime }}$, (c) $1$, $2$, $3^{^{\prime }}$ and $%
4^{^{\prime }}$, (d) $1$, $2$, $3$ and $4^{^{\prime }}$, (e) $1$, $2$, $3$
and $4$ have $1/2$ spin. The solid lines represent the $n.n.$ interactions $%
J_1$, and the dot lines the $n.n.n.$ interactions $J_2$. The cluster can be
divided into two subsystems $A$ and $B$. The solid circles denote the sites
in $A$, and the solid and open squares those in $B$.}
\label{Fig. 1}
\end{figure}

\begin{figure}[tbp]
\caption{$Q(\pi ,\pi )$ and $Q(\pi ,0)$ vs. $J_2/J_1$. Curves (a), (b), (c),
(d) and (e) are $Q(\pi ,\pi )$ for real clusters (a), (b), (c), (d) and (e)
respectively (Fig. 1). Curve $Q(\pi ,0)$ is for real cluster (a).}
\label{fig2}
\end{figure}
\begin{figure}[tbp]
\caption{Several kinds of $\overline{S_i^z}$ distributions on sites of
cluster (a) (Fig. 1). The lengthes of arrows represent the magnitudes of $%
\overline{S_i^z}$. a) $\overline{S_i^z}=$ $0.236$ for the corner sites. b) $%
\overline{S_i^z}=0.188$ for the corner sites. c) $\overline{S_i^z}=0.100$
for the corner sites. d) $\overline{S_i^z}=0.174$ for the center site.}
\label{fig3}
\end{figure}


\begin{references}
\bibitem{Gelfand}  M. P. Gelfand, R. R. P. Singh and D. A. Huse, Journal of
statistical Physics {\bf 59}, 1093 (1990); H. Q. Lin, Phys. Rev. B {\bf 42},
6561 (1991).

\bibitem{Chandra}  P. Chandra and B. Doucot, Phys. Rev. B {\bf 38}, 9335
(1988); M. Takahashi, Phys. Rev. B {\bf 40}, 2494 (1989); H. Nishimori and
Y. J. Saika, J. Phys. Soc. Jpn. {\bf 59}, 4454 (1990).

\bibitem{Chubukov}  A. V. Chubukov, S. Sachdev, and J. Ye, Phys. Rev. B {\bf %
49}, 11919 (1994).

\bibitem{Kotov}  V. N. Kotov {\it et al.}, Phys. Rev. Lett. {\bf 80}, 5790
(1998). P. V. Shevchenko, A. W. Sandvik and O. P. Sushkov, Cond-matt/9905227.

\bibitem{Johnston}  T. Nakamura and N. Hatano, J. Phys. Soc. Jpn. {\bf 62},
3062 (1993); D. C. Johnston, M. Troyer, S. Miyahara, D. Lidsky, K. Ueda, A.
Azuma, Z. Hiroi, M. Takano, M. Isobe, Y. Ueda, M. A. Korotin, V. I.
Anisimov, A. V. Mahajan and L. L. Miller, cond-mat/0001147 (2000).

\bibitem{Sandvik}  A. W. Sandvik and D. J. Scalapino, Phys. Rev. Lett. {\bf %
72}, 2777 (1994); ZhengWeihong, Phys. Rev. B {\bf 55}, 12267(1997).

\bibitem{Dagotto}  E. Dagotto and A. Moreo, Phys. Rev. Lett. {\bf 63},
2148(1989); F. Figueirido, A. Karlhede, S. Kivelson, S. Sondhi, M. Rocek and
D. S. Rokhsar, Phys. Rev. B {\bf 41}, 4619 (1989).

\bibitem{Schulz}  H. J. Schulz and T. A. L. Ziman, Europhys. Lett. 18, 355
(1992), H. J. Schulz, T. A. L. Ziman and D. Poilblanc, J. Phys. (France) I6,
675 (1996).

\bibitem{Gelfand2}  M. P. Gelfand, R. R. P. Singh and D. A. Huse, Phys. Rev.
B {\bf 40}, 10801 (1989); M. P. Gelfand, {\it ibid.} {\bf 42}, 8206 (1990).

\bibitem{Read}  N. Read and S. Sachdev, Phys. Rev. Lett. 66, 1773 (1991);
ibid. 62, 1697 (1989); G. Murthy and Sachdev, Nucl. Phys. B 344, 557 (1990).

\bibitem{Kotov2}  V. N. Kotov, J. Oitmaa, O. P. Sushkov and Zheng Weihong,
Phys. Rev. B 60, 14613 (1999).

\bibitem{Singh}  R. R. P. Singh, Zheng Weihong, C. J. Hamer and J. Oitmaa,
Phys. Rev. B 60, 7278 (1999).

\bibitem{Sachdev}  S. Sachdev and R. N. Bhatt, Phys. Rev. B 41, 9323 (1990);
A. V. Chubukov and T. Jolicoeur, Phys. Rev B 44, 12050 (1991).

\bibitem{Sushkov}  O. P. Sushkov, J. Oitmaa and Zheng Weihong,
Cond-mat/0007329.

\bibitem{Reger}  J. D. Reger and A. P. Young, Phys. Rev. B 37 (1988) 5978;
M. Gross, E. Sanchez-Velasco and E. Siggia., Phys. Rev. B, 39 (1989) 2484.;
H. Q. Ding and M. S. Makivic, Phys. Rev. Lett., 64 (1990) 1449; J. Oitmaa
and D. D. Betts, Can. J. Phys., 56 (1978) 897; R. R. P. Singh and R.
Narayanan, Phys. Rev. Lett., 65 (1990) 1072.

\bibitem{Order}  The order of infinite system is uniform. But, for real
cluster, it is nonuniform. Here, the word `order' means a kind of bulk
character of spin status.
\end{references}
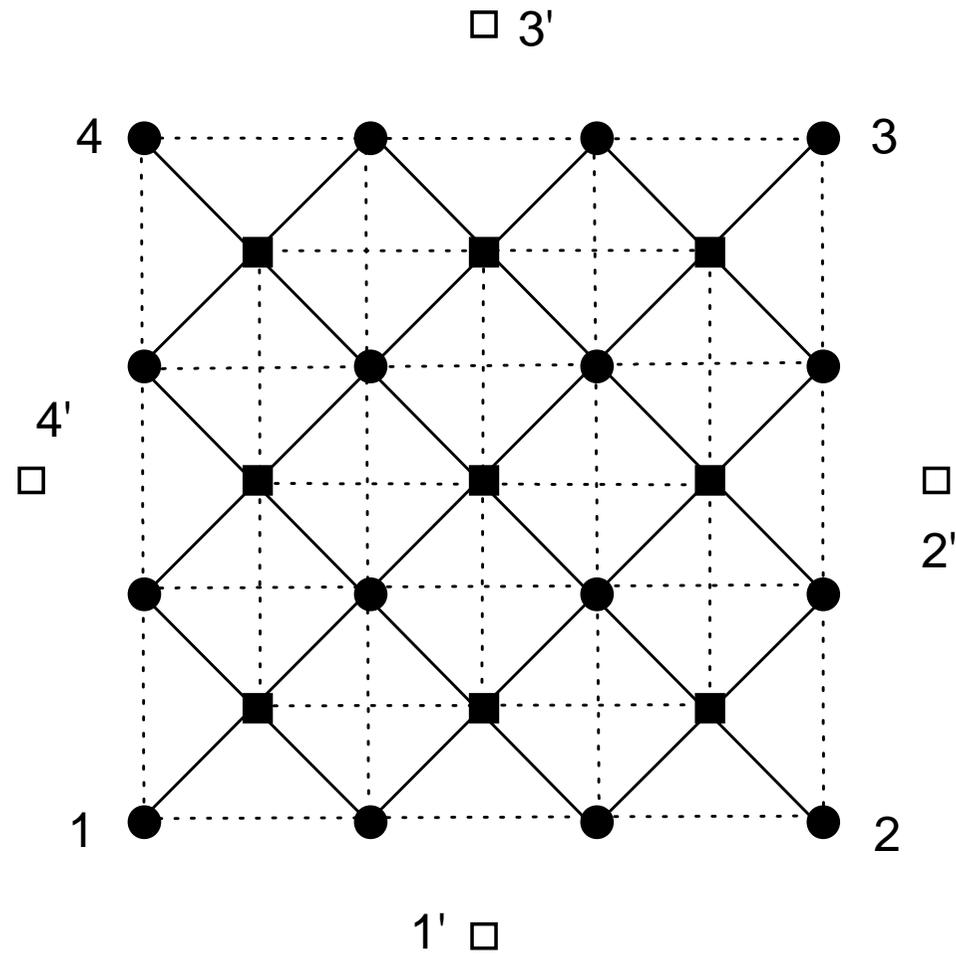
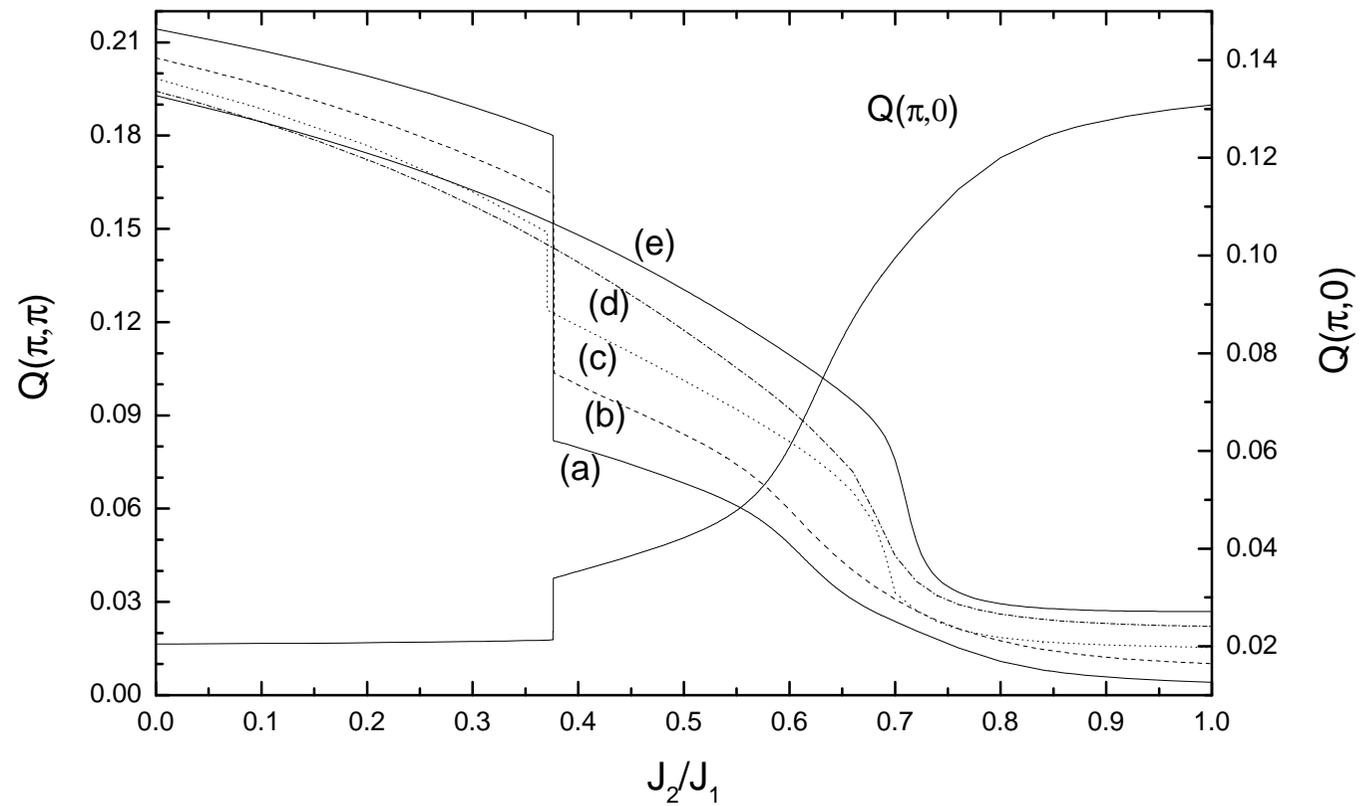
\end{document}